\documentclass[aps,pre,twocolumn,superscriptaddress,showpacs,amsmath,amssymb]{revtex4-1}

\usepackage{graphicx}% Include figure files

\newcommand{\figura}[3]{ \begin{figure}[htb] \centering \resizebox{#1}{!}{\includegraphics{#2}}\\
\caption{(color online)\label{#2} #3} \end{figure}}

\newcommand{\figuraB}[4]{ \begin{figure}[htb] \centering \resizebox{#1}{!}{\includegraphics{#2}}\resizebox{#1}{!}{\includegraphics{#3}}\\
\caption{(color online)\label{#2} #4} \end{figure}}
\newcommand{\figuraC}[5]{ \begin{figure}[htb] \centering \resizebox{#1}{!}{\includegraphics{#2}}\resizebox{#1}{!}{\includegraphics{#3}}\resizebox{#1}{!}{\includegraphics{#4}}\\
\caption{(color online)\label{#2} #5} \end{figure}}

\newcommand{\beq}[1]{\begin{equation}\label{eq:#1}}
\newcommand{\eeq}{\end{equation}}
\newcommand{\bea}{\begin{eqnarray}}
\newcommand{\eea}{\end{eqnarray}}

\newcommand{\dis}{$ \xi$}

\begin{document}

\title{
Effect of disorder on temporal fluctuations in drying induced cracking
}
\author{Gabriel Villalobos}
\email{gvillalobosc@bt.unal.edu.co}
\affiliation{Grupo de Simulaci\'on de Sistemas F\'\i sicos, Departamento de
  F\'\i sica, Universidad Nacional de
  Colombia, Crr 30 \# 45-03, Ed.~404, Of.~348, Bogota D.C., Colombia.}
\affiliation{CeiBA-Complejidad. Carrera 19A 1-37 Este   Of. C222 Bogot\'a, Colombia.}

\author{Ferenc Kun}
\affiliation{Department of Theoretical Physics. University of Debrecen, H-4010 Debrecen, P.O.Box 5, Hungary.}

\author{Jos\'e D. Mu{\~n}oz} 
\affiliation{Grupo de Simulaci\'on de Sistemas F\'\i sicos, Departamento de
  F\'\i sica, Universidad Nacional de
  Colombia, Crr 30 \# 45-03, Ed.~404, Of.~348, Bogota D.C., Colombia.}
\affiliation{CeiBA-Complejidad. Carrera 19A 1-37 Este   Of. C222 Bogot\'a, Colombia.}

\begin{abstract}
We investigate by means of computer simulations the effect of structural disorder on the statistics of cracking for a thin layer of material under uniform and isotropic drying.  For this purpose, the layer is discretized into a triangular lattice of springs with a slightly randomized arrangement. The drying process is captured by reducing the natural length of all springs by the same factor, and the amount of quenched disorder is controlled by varying the width $\xi$ of the distribution of the random breaking thresholds for the springs. Once a spring breaks, the redistribution of the load may trigger an avalanche of breaks, not necessarily as part of the same crack.  Our computer simulations revealed that the system exhibits a phase transition with the amount of disorder as control parameter: at low disorders, the breaking process is dominated by a macroscopic crack at the beginning, and the size distribution of the subsequent breaking avalanches shows an exponential form. At high disorders, the fracturing proceeds in small-sized avalanches with an exponential distribution, generating a large number of micro-cracks which eventually merge and break the layer. Between both phases a sharp transition occurs at a critical amount of disorder $\xi_c = 0.40\pm 0.01$, where the avalanche size distribution becomes a power law with exponent $\tau=2.6\pm 0.08$, in agreement with the mean-field value $\tau=5/2$ of the fiber bundle model. Moreover, good quality data collapses from the finite-size scaling analysis show that the average value of the largest burst $\left< \Delta_{max}\right>$ can be identified as the order parameter, with $\beta/\nu = 1.4$ and $1/\nu \simeq 1.0$, and that the average ratio $\left<m_2/m_1\right>$ of the second $m_2$ and first moments $m_1$ of the avalanche size distribution shows similar behaviour to the susceptibility of a continuous transition, with $\gamma / \nu = 1.$, $1 / \nu \simeq 0.9$. These results suggest that the disorder induced transition of the breakup of thin layers is analogous to a continuous phase transition.  \end{abstract}


\maketitle

 \section{Introduction}

Desiccation induced cracking of a thin layer of materials is an interesting scientific problem with a broad spectrum of technological applications.  Painted surfaces, thin wood layers, antirreflective and UV-protecting coatings on glasses and thin-film manufacturing processes like chemical-bath and sol-gel depositions are just a few of a large set of examples where dessication-induced cracks might be avoided.  In contrast with fractures induced by collisions or tensile stresses, drying of a thin-layer material is a homogeneous and isotropic process occurring everywhere inside the layer, whose statistical behavior could be different from those previously studied cases.

During the past decades experiments have revealed that the competition of crack formation inside the layer with the delamination from the substrate leads to a gradual breakup into pieces of polygonal shapes whose characteristic length scale is determined by the layer's thickness.  Recently Nakahara and Matsuo addressed the possibility of controlling the structure and time evolution of crack patters in pastes, i.e.\ dense colloidal suspensions, by subjecting the paste to vibration or flow before the onset of drying \cite{NAKAHARAMATSUO07,PhysRevE.74.045102,1742-5468-2006-07-P07016}.  Detailed experiments showed that those mechanical excitations induce structural rearrangements and plastic deformations getting imprinted onto the patterns of the drying crack network. A large number of experiments have been performed to understand the microscopic mechanism leading this memory effect.  They conclude that the interaction (attractive or repulsive) among the colloidal particles and the amount of disorder (shape regularity) govern the pattern formation \cite{2011arXiv1101.0953M}. A very interesting effect of the amount of disorder has recently been pointed out in other types of fracture processes, as well. Fiber bundle models with equal load sharing exhibit a power-law distribution of the bursts' sizes. \cite{alava-2006-55,PradhanHansenEtAlFailure09,hidalgo_pre_2009}. Mixing fibers with widely different breaking strengths reveals a phase transition between two regimes with different power law exponents of the bursts' distribution. The transition takes place at a well defined system composition, and it has shown to be analogous to a continuous phase transition \cite{hidalgo_universality_2008}.
 
In the present paper we focus on the effect of the amount of quenched disorder on the process of gradual breakup induced by desiccation, motivated by the works of Nakahara and Matsuo \cite{NAKAHARAMATSUO07,PhysRevE.74.045102,1742-5468-2006-07-P07016} and Kitsunezaki \cite{PhysRevE.60.6449}.  The thin layer of paste is discretized into a regular triangular lattice of springs with fixed ends at the boundary. Drying is captured by gradually decreasing the natural length of all springs, inducing unbalanced forces and leading to the final breaking of some bonds. The quenched disorder of the material is represented by a random distribution of the bonds' breaking thresholds, and the the amount of disorder is given by distribution's width.  There are numerous studies in the literature on the spatial arrangement of cracks \cite{tarafdar_jpcdm2010,tarafdar_physa2011,0953-8984-19-35-356206,PhysRevE.60.6449}.  Hereby, we focus on the temporal evolution of the crack by analyzing the statistics of avalanches of breaking bonds. A quasi-static desiccation process is implemented by shrinking all the springs by the same rate until a first bond breaks. This induces a redistribution of loads that can be followed by an avalanche of breaks. Simulations revealed that the layer's breakup proceeds in bursts whose dynamic and statistical features strongly depend on the amount of disorder: At low disorders a dominating crack nucleates, creating an extended free surface inside the system. As a consequence, the size distribution of bursts is discontinuous, i.e.\ small sized avalanches have an exponential distribution while larger ones show a peak distribution, with a gap in between.  At high disorders, in contrast, the breakup process consists of a large number of small-sized avalanches with an exponential distribution. Varying the amount of disorder a transition occurs between the two phases at a critical disorder where the size distribution becomes a power law. Analyzing the finite size scaling behavior of the largest and average burst sizes we demonstrate that the disorder-induced transition of our system is analogous to a continuous phase transitions.

\section{Model construction}
The thin layer of material to be dried is discretized into a uniform two-dimensional triangular lattice with a slightly distorted arrangement. The nodes represent point masses and the bonds among them provide for cohesive forces. Nodes at the boundaries are set fixed, while internal nodes can move in the two-dimensional plane.  System size $L$ is defined as the number of nodes in the lower boundary row of the lattice. Simulations were carried out for the values $L=16,24,30,36,42,50$. The bonds are simple central force springs, each one with a different spring constant $k=El/A$, where $l$ is the natural length of the spring, $A$ is a uniform cross section and $E$ denotes for the Young modulus of the layer's material.

The effect of desiccation is implemented by reducing the natural length of all springs by the same ratio. It is controlled by an increasing parameter $\alpha$, given by
 \beq{alpha} 
 \alpha = \frac{l_0-l}{l_0}, 
 \eeq 
where $l$ is the current natural length of the spring and $l_0$ denotes its initial value, computed as the initial distance between the two nodes connected by the spring. Springs are assumed to have a finite strength: When the load $\sigma$ on a spring exceeds its breaking threshold $\sigma_{th}^i$ ($i=1,\ldots , N$), the spring breaks instantaneously and it will never be restored (no healing). The quasi-static process of a very slow drying is assured by looking for the smaller $\alpha$ to break a spring.

\figuraB{4.1cm}{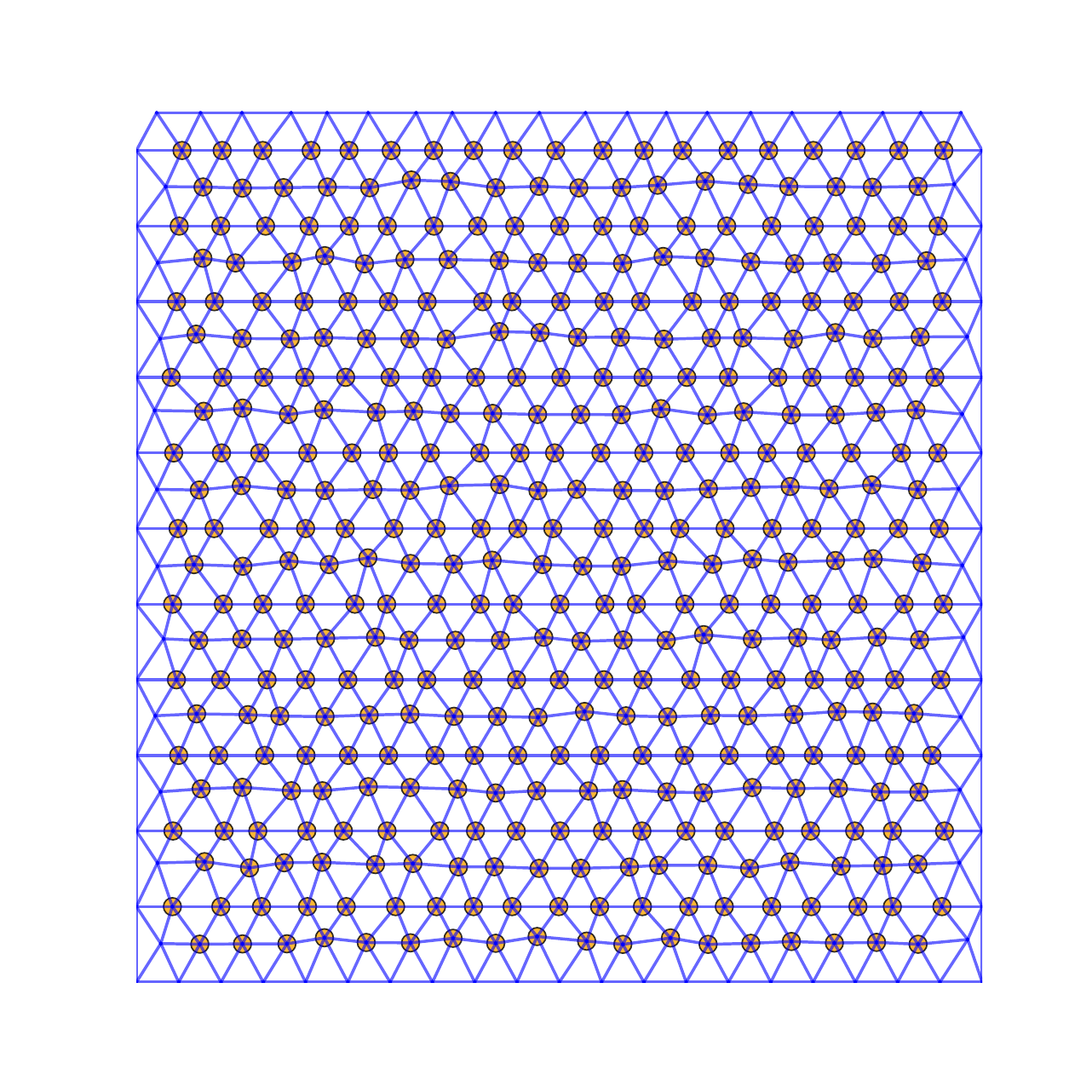}{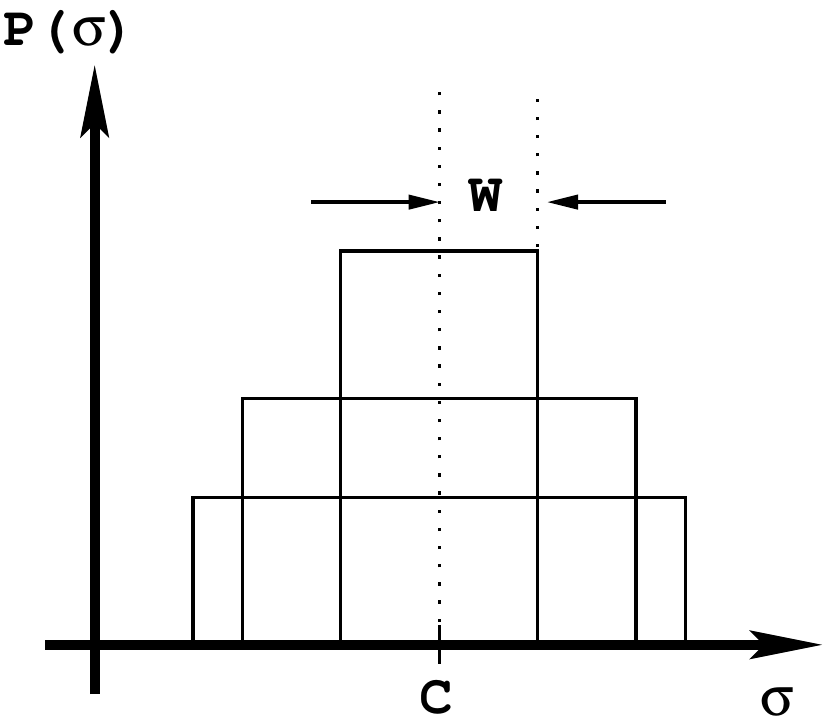}{\emph{Left:}
  Example of initial randomized geometry. \emph{Right:}
Uniformly distributed breaking thresholds are centered at $C=1$
with half-width $W$. The amount of strength disorder is controlled by varying
the value of $W$ between 0 and 1.}
 
The quenched disorder of the material is represented by the randomness of the breaking thresholds $\sigma_{th}$.  In order to be able to vary the amount of heterogeneity in the system, we consider a uniform distribution of threshold values centered at $C=1$ with half-width $0\leq W \leq 1$, that is with probability density \begin{eqnarray}
 p(\sigma_{th}) = \frac{1}{2W}, \ \ \ \mbox{for} \ \ \ C-W\leq \sigma_{th} \leq C+W.
\end{eqnarray}
Hence, the strength of disorder is characterized by the dimensionless parameter
\begin{eqnarray}
  \xi = \frac{W}{C},
\end{eqnarray}
ranging from 0 (no disorder) to 1 (highest disorder). We note that there is also a slight amount of structural disorder in the system: Even though the spring lattice is almost a triangular regular grid with spacing $a$, each node is randomly set inside a circle of radius $0.1a$, centered at the regular place.  This slight distortion is sufficient to prevent the formation of artificial crack patterns, but otherwise, it does not have any relevant role in the breakup process.
 
After removing a spring, the system is relaxed by numerically integrating through the Verlet algorithm the equation of motion for each node, 
\begin{eqnarray}
  m\ddot{\vec{r}}_i = \vec{F}_i -c\vec{v}, \qquad i=1,\ldots , N
\end{eqnarray}
where a velocity-dependent damping force (with damping coefficient $c=10$) has
been added to assure for relaxation. The node mass $m$ is proportional to the
area of the set of all points closer to this node than to the other ones (that
is the Voronoi polygon of the dual lattice for this geometry).  In the present
paper, because the geometry is close to a regular triangular lattice,
$m\approx 0.86$ for all internal nodes \cite{triangle}. The relaxation causes
the redistribution of the load dropped by the broken bond over the remaining
ones, leading to spatial correlations in the system. It has to be emphasized
that this force rearrangement itself may result into subsequent breakings, as
more springs surpass their breaking thresholds while keeping their relaxed
length fixed. As a consequence, a single spring breaking induced by drying may
trigger an avalanche of breakings.  The avalanche size $\Delta$ is defined as
the number of broken bonds during the avalanche, including also the first one
whose breaking was provoked by the drying step.  Thus \emph{cracks}, defined
as clusters of contiguous broken bonds, appear.  Due to the long-range load redistribution, a single avalanche can generate several cracks, and it is also possible that several independent avalanches contribute to a given crack.

In the framework of the model, the desiccation is treated as a stepwise process, where the number of shrinkage steps provides a measure of time $t$.  During drying, the amount of shrinkage is characterized by the relative length reduction $\alpha = \Delta l/l_0$ of spring elements, so that we use $\alpha$ as the degree of desiccation.  The initial value of $\alpha$ at the beginning of the loading process is $\alpha_0=0$. The drying process ends when the humidity content on the sample drops to $0$. We consider this to be equivalent to a finite value of the shrinkage parameter $\alpha_f$, which in the present model is set to the arbitrary value of $\alpha_f=0.3$. A large amount of computer simulations were performed in order to understand the effect of the strength of disorder on the breakup process of thin layers of heterogeneous materials. Specifically,  $N=1000$ for $L=16$, $L=24$ and $L=30$, $N=800$ for $L=36$, $N=500$ for $L=42$ and $N=50$ for $L=50$, systems were generated for each value of the amount of disorder $\xi$ by changing the seed of the random number generator. 

In the present paper we focus on the effect of the amount of disorder on the statistics of avalanches.

\section{Breakup Process}
As explained before, the desiccation-induced breakup of our discrete model is not a smooth process, but evolves in bursts (avalanches).  The dessication parameter $\alpha$ grows until a single spring brakes, triggering the burst and leading to load redistributions and additional rearrangements that eventually brake more springs. The burst stops when no more springs brake with that dessication factor.  Both these temporal fluctuations of the bursting activity and the emerging crack structure are strongly affected by the degree of disorder in the system. To give a quantitative characterization of the effect of disorder on the breakup process, simulations were carried out by varying the value of $\xi$ in a broad range.  \figuraC{2.8cm}{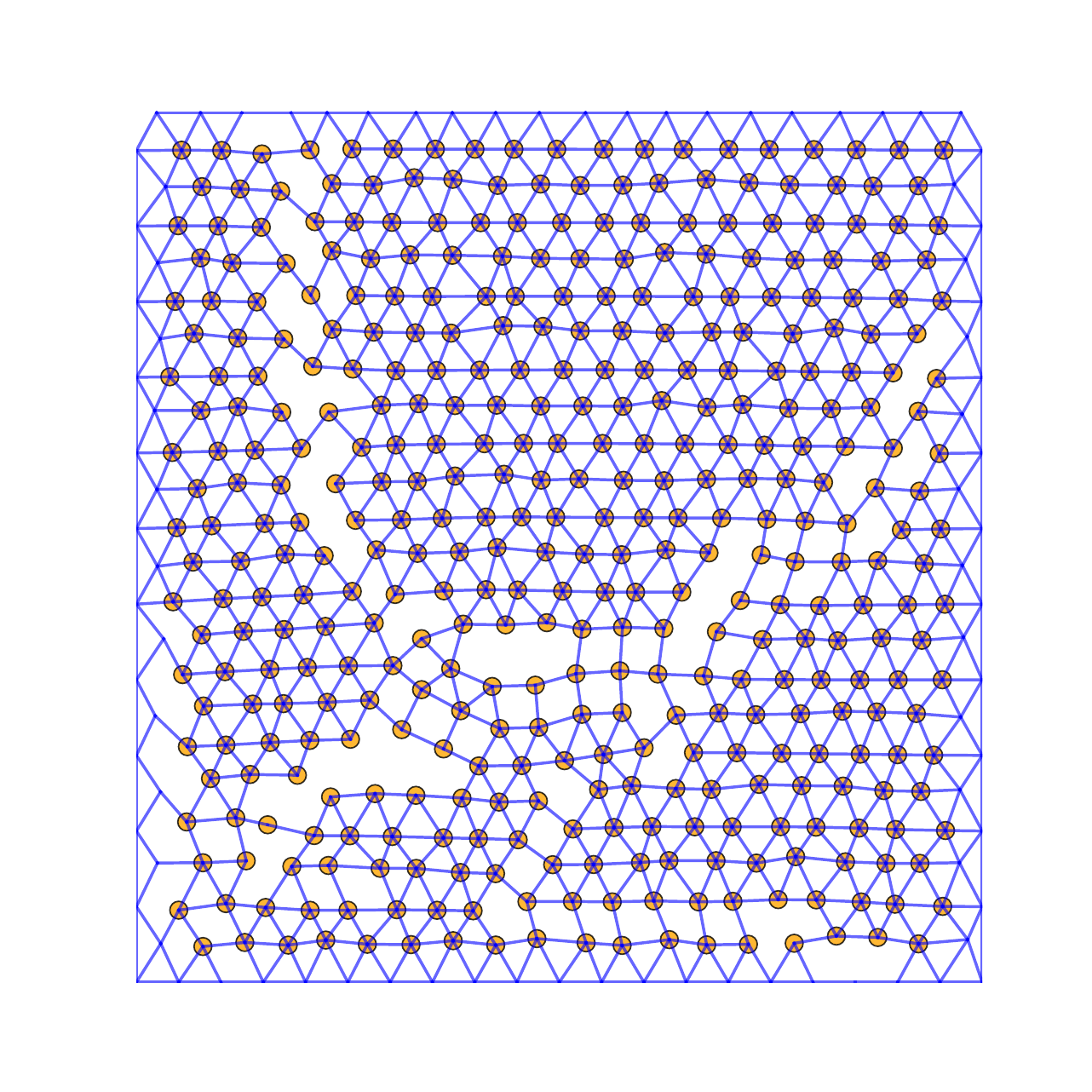}{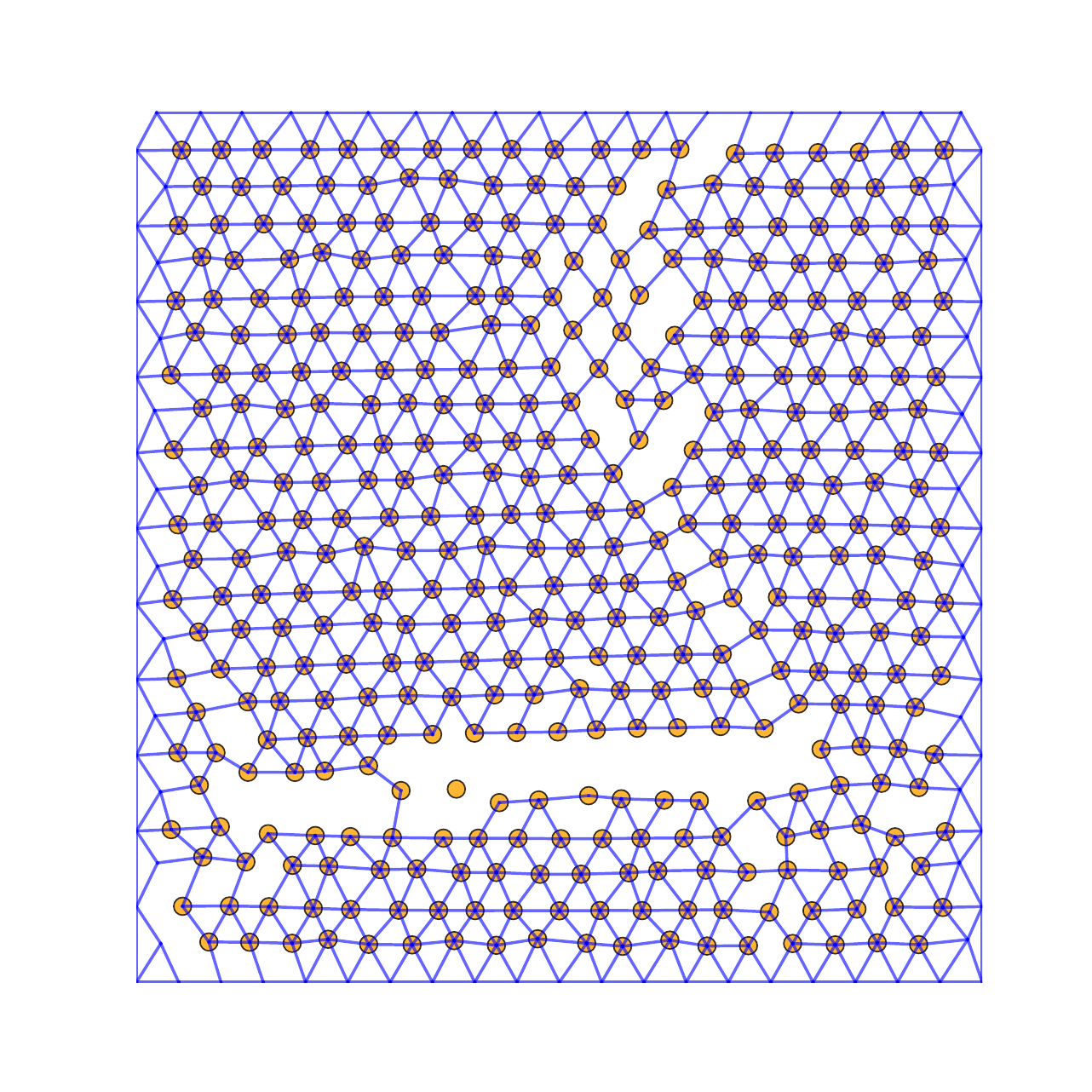}{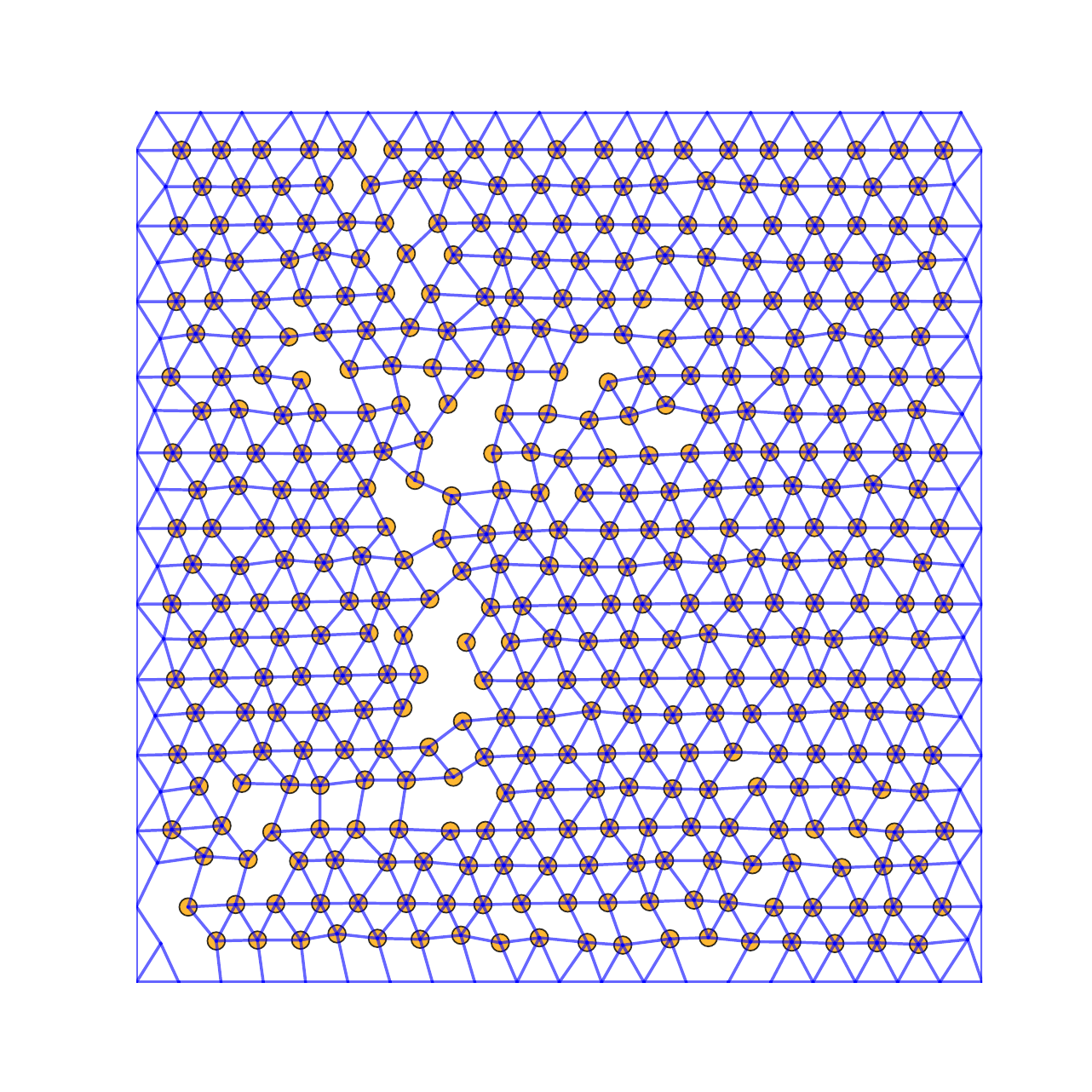}{First
  large avalanche for three different widths of disorder, showing how
  the narrower breaking distributions favor the appearance of large
  fractures.  \emph{Left:
  }\dis$=0.01$,
  \emph{Center:}
  \dis$=0.1$,
  \emph{Right:}.\dis$=0.3$.
  \cite{VILLALOBOS2011VIDEOS}.
}

Large avalanches consist of spatially correlated bonds being broken; even though not necessarily as a single expanding crack of contiguous bonds. Such spatial correlation takes place at the beginning of the breaking process for low amounts of disorder in the distribution of breaking thresholds, where large avalanches appear early on (as seen in Fig.\ \ref{fig2a.pdf}), due to stress accumulations.  Conversely, large amounts of disorder do not generate correlated crack growths during the early stages of the evolution. The initial breakings tend to be random, scattered all over the lattice, with small avalanches of size $\Delta \simeq 1$. Only after a fraction of bonds breaks and small micro-cracks have appeared in the system, these micro-cracks can coalesce by means of \emph{gaping cracks,} that are clusters of broken bonds joining two micro-cracks. As drying continues, both for small and large disorder, the newly created free surfaces hinder the appearance of large avalanches by relaxing the stress in the remaining intact parts of the spring network.  The overall shape of cracks is similar to what has been observed for drying processes in composite materials, where the disorder comes from random distributions of the three types of bonds representing the interaction between the ingredients of a composite (see \cite{tarafdar_jpcdm2010,tarafdar_physa2011,0953-8984-19-35-356206}).

\figura{9cm}{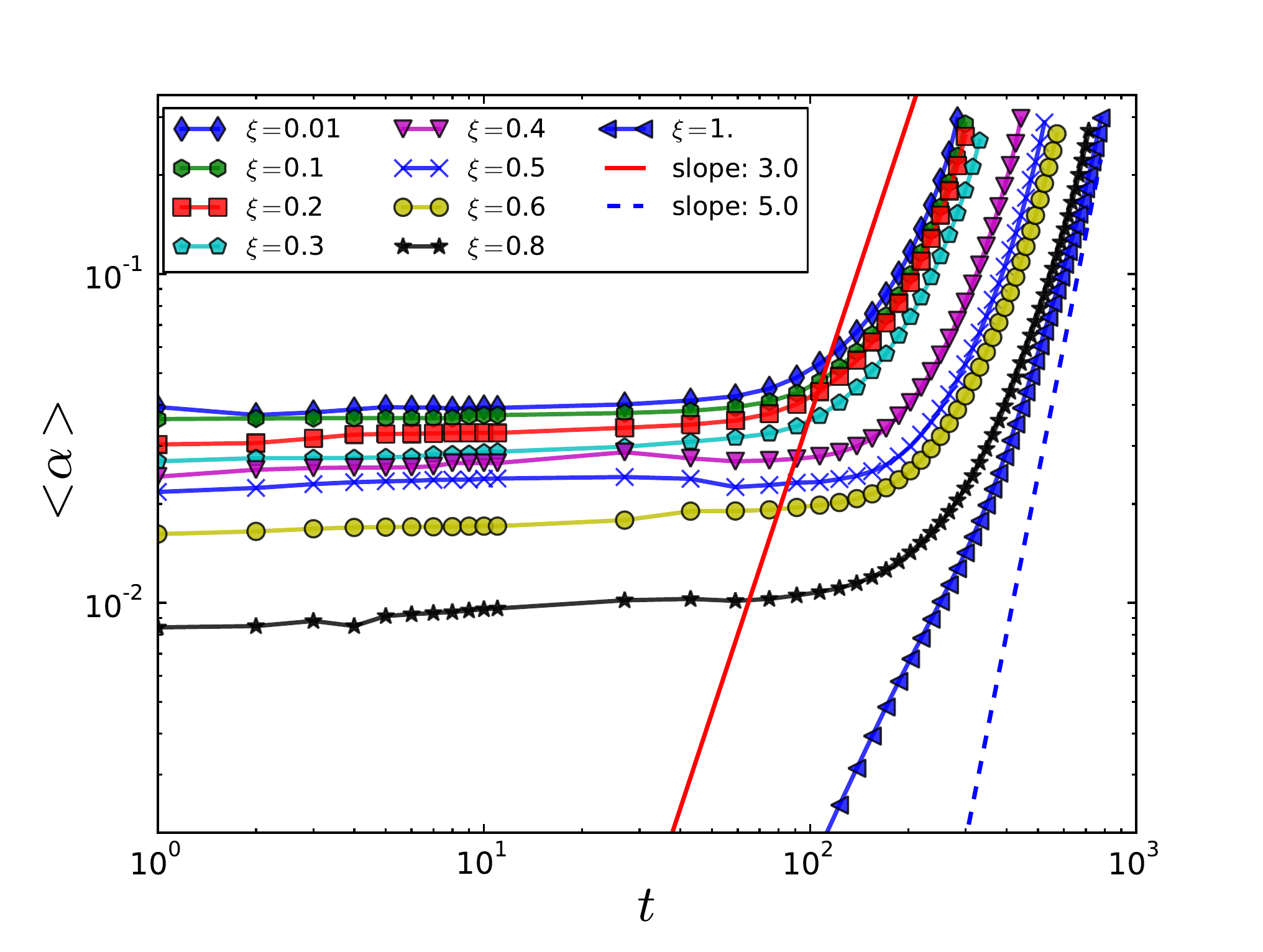}{The relative length
  reduction $\left<\alpha\right>$ of springs as a function of time for
  several values of the amount of disorder $\xi$. The system size was
  set to $L=30$. Data values are binned with a bin size $\Delta t=1$ and $\Delta t=16$ for $0<t<10$
and for $t>10$, respectively.}  

To follow the time evolution of the process, we evaluated the average value of the desiccation parameter $\left<\alpha\right>$ as a function of time $t$. It is important to emphasize that the curves of $\left<\alpha\right>(t)$ presented for several values of $\xi$ in Fig.\ \ref{fig3.pdf} have two distinct regimes: at the beginning of the drying process the value of $\left<\alpha\right>$ remains nearly constant over two orders of magnitude in time, which implies a large number of shrinking steps with tiny length changes $\delta\alpha\ll 1$. This slow regime is then followed by a sudden acceleration of the process where springs suffer a considerable length reduction. The reason of the crossover from slow to rapid shrinkage is the formation of a dominating crack spanning the system in one or both directions. Due to the large free surface inside the lattice, a larger reduction in the length of all springs is required to trigger the next breaking event. The analysis of the spatial structure of damage shows that the transition point can be identified as the exact time when the spanning crack appears. This occurs for all levels of disorder $\xi$; nevertheless, the geometrical structure strongly depends on the precise value of $\xi$.

It can be seen in Fig.\ \ref{fig3.pdf} 
that the curves of $\left<\alpha\right>(t)$ shift to the
right with increasing values of $\xi$. It means that the 
the total duration $t_f$ of the process, i.e.\ the total number
of shrinkage steps required to reach some value $\alpha_f$, is an increasing
function of the amount of disorder, as illustrated in Fig.\ 
\ref{fig4.pdf}.
\figura{8.8cm}{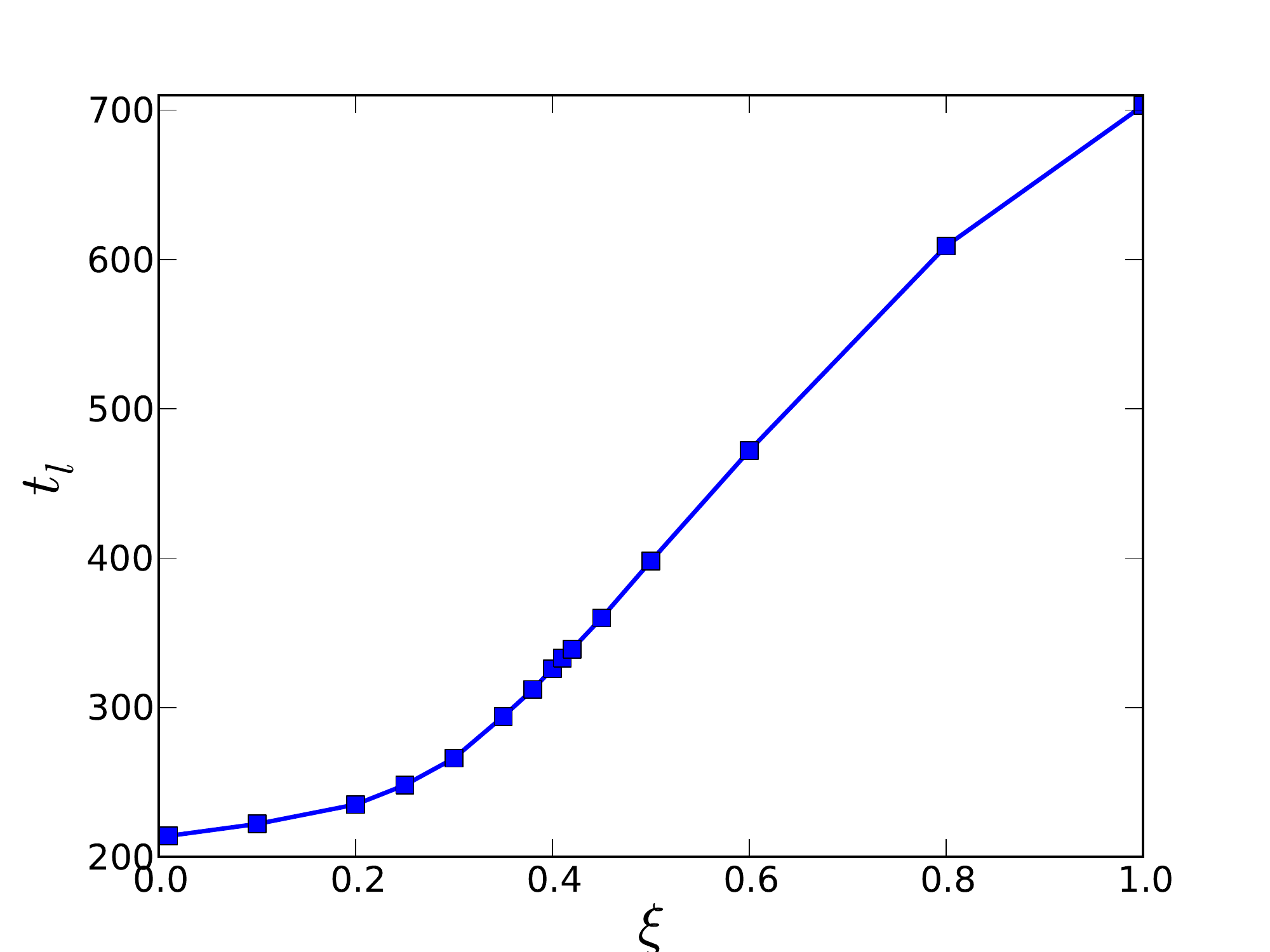}{
  Average number of steps $t_f$ needed to reach the shrinkage value of
  $\alpha_f=0.7$, for different values of the disorder \dis, for a
  system of size $L=30$.}
In other words, the process slows down with increasing disorder, i.~e. a larger number of smaller steps are required to drive the fracture process. It can be observed in Fig.\ \ref{fig4.pdf} that the monotonically increasing $t_f$ curve has a curvature change around $\xi_c\approx 0.5$, where the slope of the function has a maximum. Looking at the rate of change, the system seems to have two phases depending on the amount of disorder: a rapidly evolving low disorder phase and a slowly proceeding high disorder one, separated by the inflection point.  Both quantities we have considered until now are global characteristics of the lattice breaking process. In the following we analyze the statistical features of the individual avalanches of spring breakings.

\section{Avalanche dynamics}

Let us start by analyzing $\overline \Delta $, defined as the average size of avalanches decreased by the minimum avalanche size, $\Delta_{min}=1$, as a function of $\alpha$.

\figura{8.8cm}{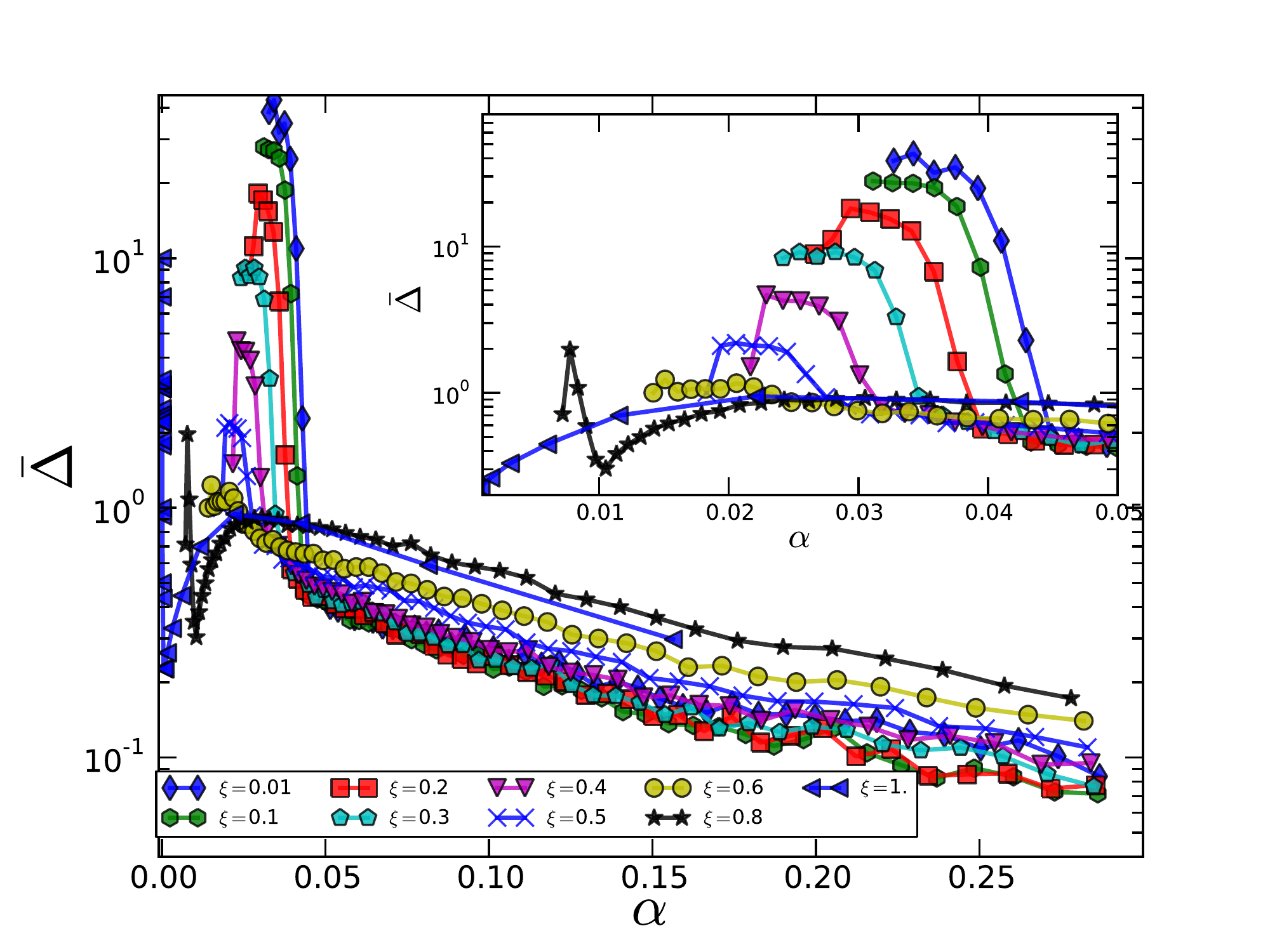}{Arithmetic average of avalanche sizes $\overline \Delta(\alpha)$ as function of the shrinkage $\alpha$ for different values of disorder \dis. Each curve is the average over several instances of the system. \emph{Inset} For very high disorders (for example \dis$=1.0$), there is a single maximum, corresponding to a single large avalanche, just at the start. This first maximum disappears close to the transition value of disorder $\xi_c=0.6$ and, thereafter, a second local maximum appears. For all cases, there is an exponential relationship between $\overline \Delta $ an $\alpha$ at the end of the process.}
It can be observed in Fig.\ \ref{fig5.pdf} that for systems with very high disorder, $\overline{\Delta}(\alpha)$ jumps discontinuously from zero to an isolated maximum, which in turn is orders of magnitude larger than the average of the function over the whole loading process. This corresponds to a large initial avalanche, caused by the release of the strain accumulation in the system.  The hight of this maximum decreases with increasing disorder and disappears for $\xi \approx 0.7$, where $\overline{\Delta}(\alpha)$ becomes a monotonically decreasing function. This is in agreement with previous results on the geometrical structure of the crack patters, showing that for large disorders the largest crack appears by coalescence, not by a single large avalanche \cite{tarafdar_jpcdm2010,tarafdar_physa2011}.  Also notice that for smaller amounts of disorder (for $\xi \le 0.6$) a local maximum appears. This implies that the large initial crack does not completely undermine the appearance
of large avalanches; rather, some new cracks can appear at this second
value of shrinkage.  It is interesting to note that in Fig.\ \ref{fig5.pdf} straight lines are obtained on a semi-logarithmic plot, evidencing that, for large $\alpha$ values, $\overline{\Delta}(\alpha)$ shows an exponential decay 
\beq{expdecay}
\overline{\Delta} \approx e^{-A\alpha},
\eeq
\noindent 
for all amounts of disorder. The multiplication factor $A$ is practically independent of
the value of $\xi$.

\figura{8.8cm}{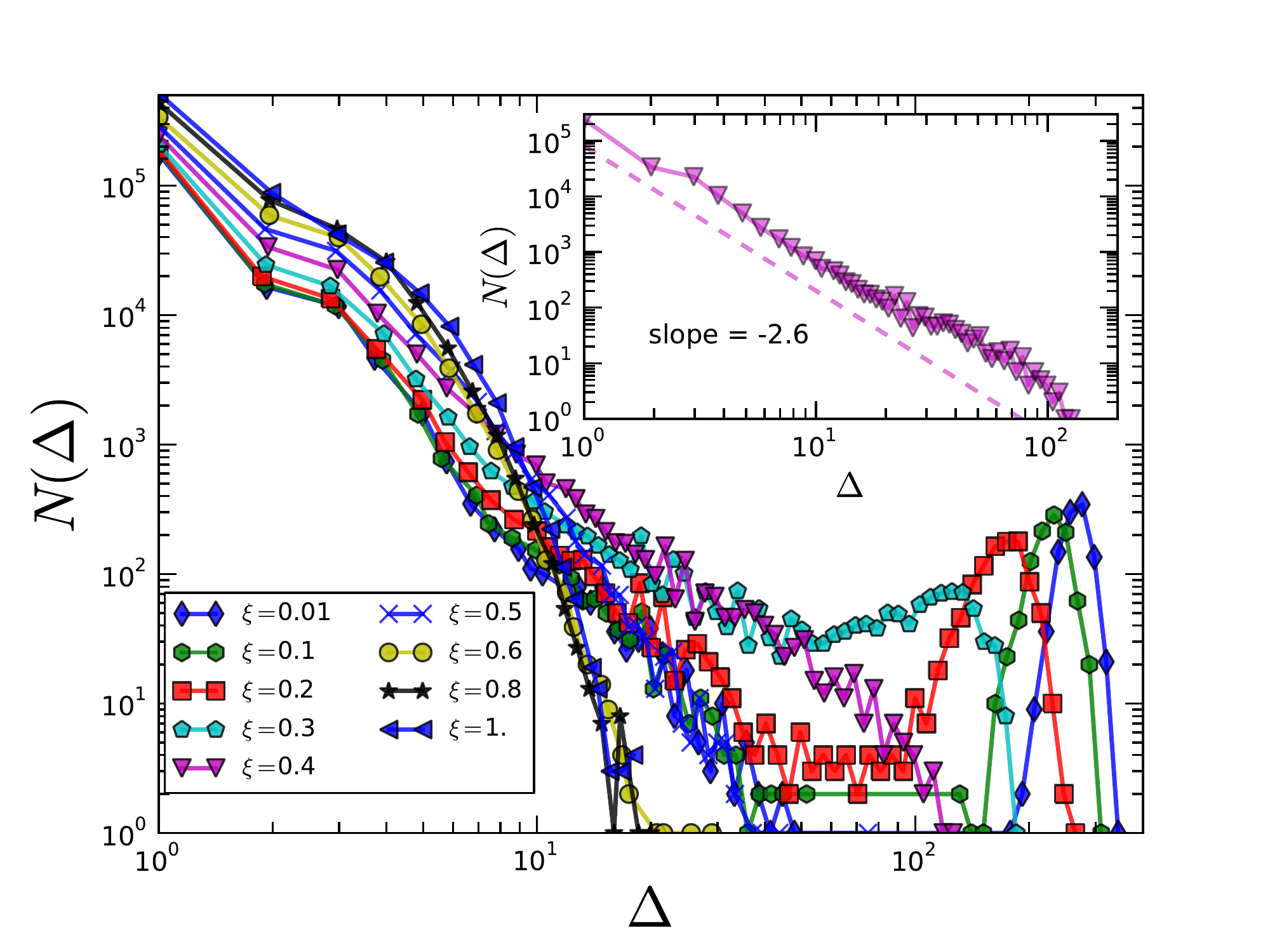}{Avalanche size
  distributions $N(\Delta) \propto P(\Delta)$ for a system of linear
  extension $L=30$, for several values of disorder \dis. {\it Inset}
  avalanche size distribution at the critical point $\xi_c$, where a
  good quality power law is obtained.}
  
The distribution of avalanche sizes $N(\Delta)$, proportional to the probability distribution of avalanche sizes $P(\Delta)$, is presented in Fig.\ \ref{fig6.pdf} for several values of the amount of disorder \dis.  It can be observed that for small disorders the distribution $P(\Delta)$ clearly separates into two distinct regimes with different functional forms: for small-sized avalanches (say, $\Delta<10$) the distribution $P$ shows an exponential decay, while for larger ones it shows a well defined peak, corresponding to a single major avalanche at the beginning of the drying (depicted in Fig.\ \ref{fig2a.pdf}).  This rapid exponential decay of small-sized avalanches is due to the appearance of a large dominating crack of almost the same as the linear extension of the system, which creates a free surface that allows for the relaxation of stresses.  The drying process later on cannot generate long breaking sequences, since nodes can easily rearrange to compensate for unbalanced forces. At high amount of disorder, in contrast, the presence of strong springs prevents the formation of a dominating crack at the onset of breaking.  The breakup proceeds in short avalanches resulting in a large number of small-sized cracks which eventually merge and relax all unbalanced forces.  Consequently, the distribution of avalanches sizes $P(\Delta)$ spans again a limited range and exhibits an exponential decay.

These two different behaviors define the two phases of the system: a first \emph{isolated large crack phase} where the fracture growth is dominated by the appearance of a macroscopic crack at the beginning of the process, almost spanning across the sample; and a second \emph{coalescence phase}, where the large fracture is formed by merging small cracks at the end of the breakup process.
The transition between the two phases is controlled by the amount of
disorder in the system. At the critical value $\xi_c =
0.40+\pm 0.01$, the probability distribution becomes a power law 
\begin{eqnarray}
  P(\Delta) \sim \Delta^{-\tau} \quad ,
\end{eqnarray}
with no characteristic length scale (see the inset of Fig.\ \ref{fig6.pdf}).
The value of the exponent $\tau$ was determined numerically as $\tau=2.6\pm0.08$.
It is interesting to note that this value falls very close to the 
mean-field exponent $\tau=5/2$ of fiber bundles 
\cite{PradhanHansenEtAlFailure09,PhysRevE.74.016122,hidalgo_pre_2009},
supporting the idea that the statistics of avalanches in drying induced cracking 
is dominated by the long range
nature of the load redistributions.

\figura{8.8cm}{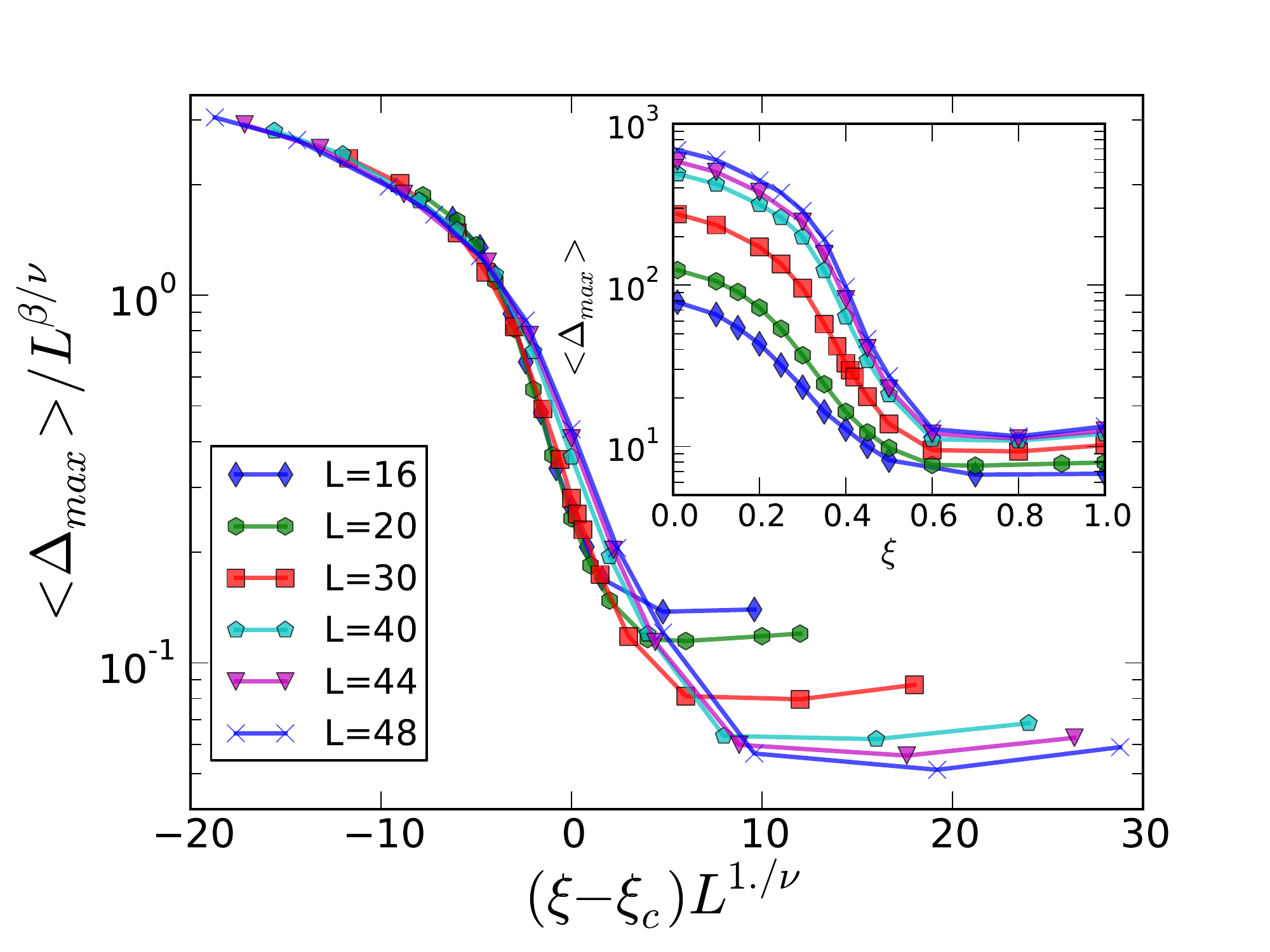}{Arithmetic average of the largest
  avalanche, for different system sizes, as a function of the disorder
  of the system \dis. The different data curves collapse in a single
  shape when the transformation is applied, implying: $\xi_c=0.4$, $\beta / \nu = 1.4$ and $1 / \nu = 1.$ \emph{Inset}
  Original data without rescaling.}

In order to understand the nature of this disorder-induced transition we computed the average value of the largest burst $\left< \Delta_{max}\right>$, which can be considered as the order parameter of the transition. Indeed, it can be observed in Fig.\ \ref{fig7.pdf} that $\left< \Delta_{max}\right>$ is a monotonically decreasing function of the amount of disorder $\xi$, with an inflection point at the transition ($\xi_c\approx 0.6$ for the system size $L=48$).  Increasing the system size $L$ shift the curves of $\left< \Delta_{max}\right>$ towards higher disorder values, but the functional form remains the same. Actually, Fig.\ \ref{fig7.pdf} shows that all curves for different system sizes $L$ can be collapsed onto a universal master curve by appropriately rescaling both axis on the plot. The good quality data collapse implies the scaling structure
\beq{collapseorder}
\left< \Delta_{max}\right>(L,\xi) = L^{\beta/\nu} f((\xi-\xi_c)L^{1/\nu}),
\eeq
where the value of the critical exponents were obtained numerically as
$\beta/\nu = 1.4$ and $1/\nu=1.0$.

\figura{8.8cm}{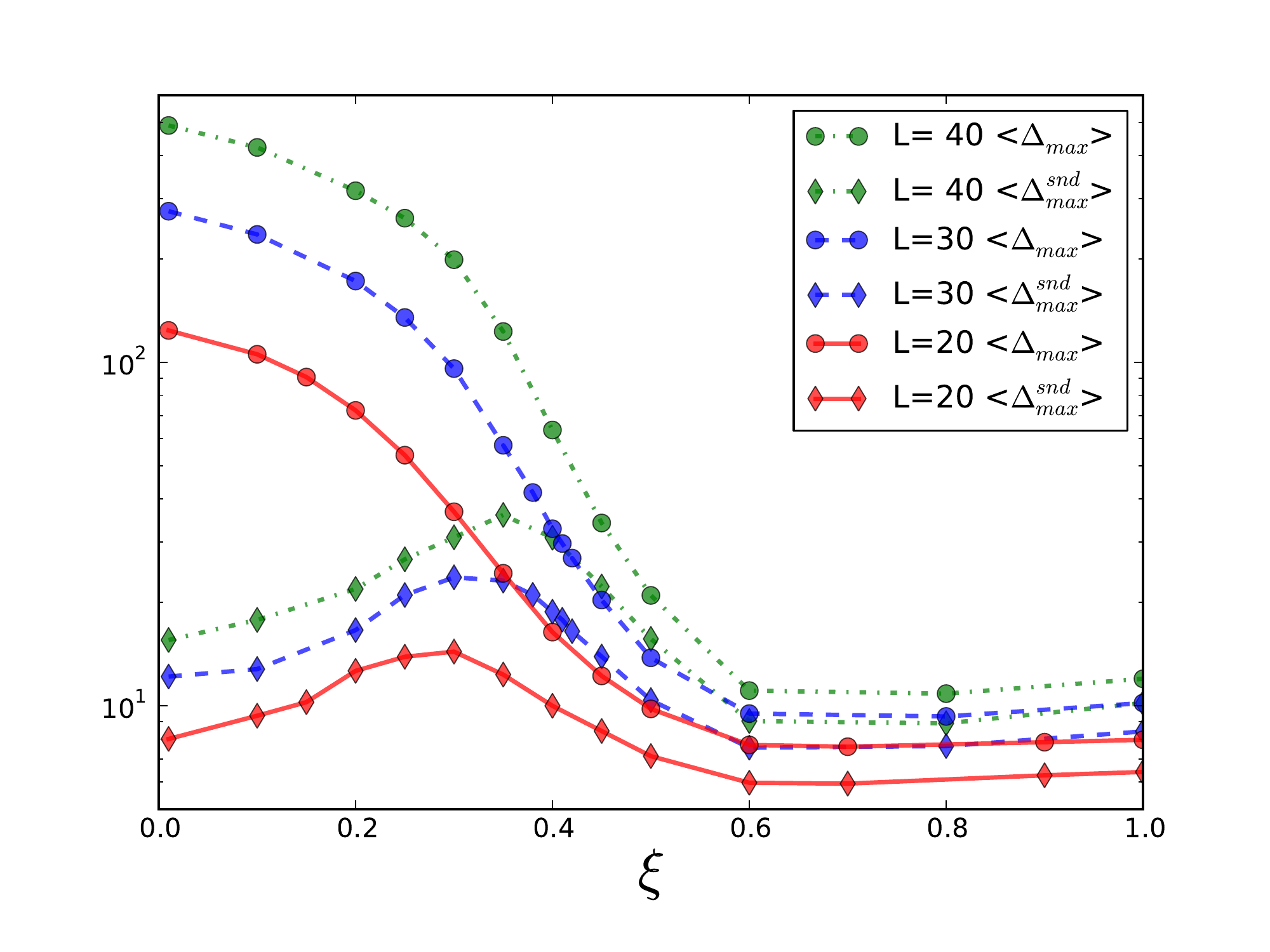}{ Arithmetic average of the largest $\left< \Delta_{max}\right>$ and second largest $\left< \Delta^{2nd}_{max}\right>$ avalanches for different system sizes, as a function of the disorder \dis.}

Fig.\ \ref{fig8.pdf} compares the average of the largest $\left< \Delta_{max}\right>$ and second largest $\left< \Delta^{2nd}_{max}\right>$ bursts.  It can be observed that below the critical point $\xi < \xi_c$ the size of the largest avalanche is a finite fraction of the total number of bonds in the lattice, while the second largest burst is orders of magnitude smaller; in contrast, in the high-disorder regime $\xi > \xi_c$ the largest and second largest bursts are similar and decrease together as $\xi$ increases. Based on this behavior, let us define the average burst size $\left<\Delta\right>$ as the average value of the ratio of the second $m_2$ and first moments $m_1$ of the avalanche size distribution \cite{stauffer_percolation},
\begin{eqnarray}
  \left<\Delta\right> = \left< \frac{m_2}{m_1}\right> \quad,
\end{eqnarray}
where the $k$th moment of the burst sizes is defined as
\begin{eqnarray}
m_k = \sum_{i=1}^{K} \Delta^k -\Delta_{max}^k \quad.
\end{eqnarray}
When plotted against $\xi$, this average burst size exhibits the remarkable feature of showing a maximum at the same critical disorder $\xi_c$ obtained in the previous analysis of the largest bursts (Fig.\ \ref{fig9.pdf}).  Both results demonstrate that the interplay between strength disorder and stress inhomogeneities around cracks leads to the emergence of a critical state of the system where the bursting activity becomes scale-free. Note that in the evaluation of the average burst size $\left<\Delta\right>$ the largest burst is always omitted; thus, the maximum in Fig.\ \ref{fig9.pdf} implies that the characteristic avalanche size becomes comparable to the system size at the critical disorder.
\figura{8.8cm}{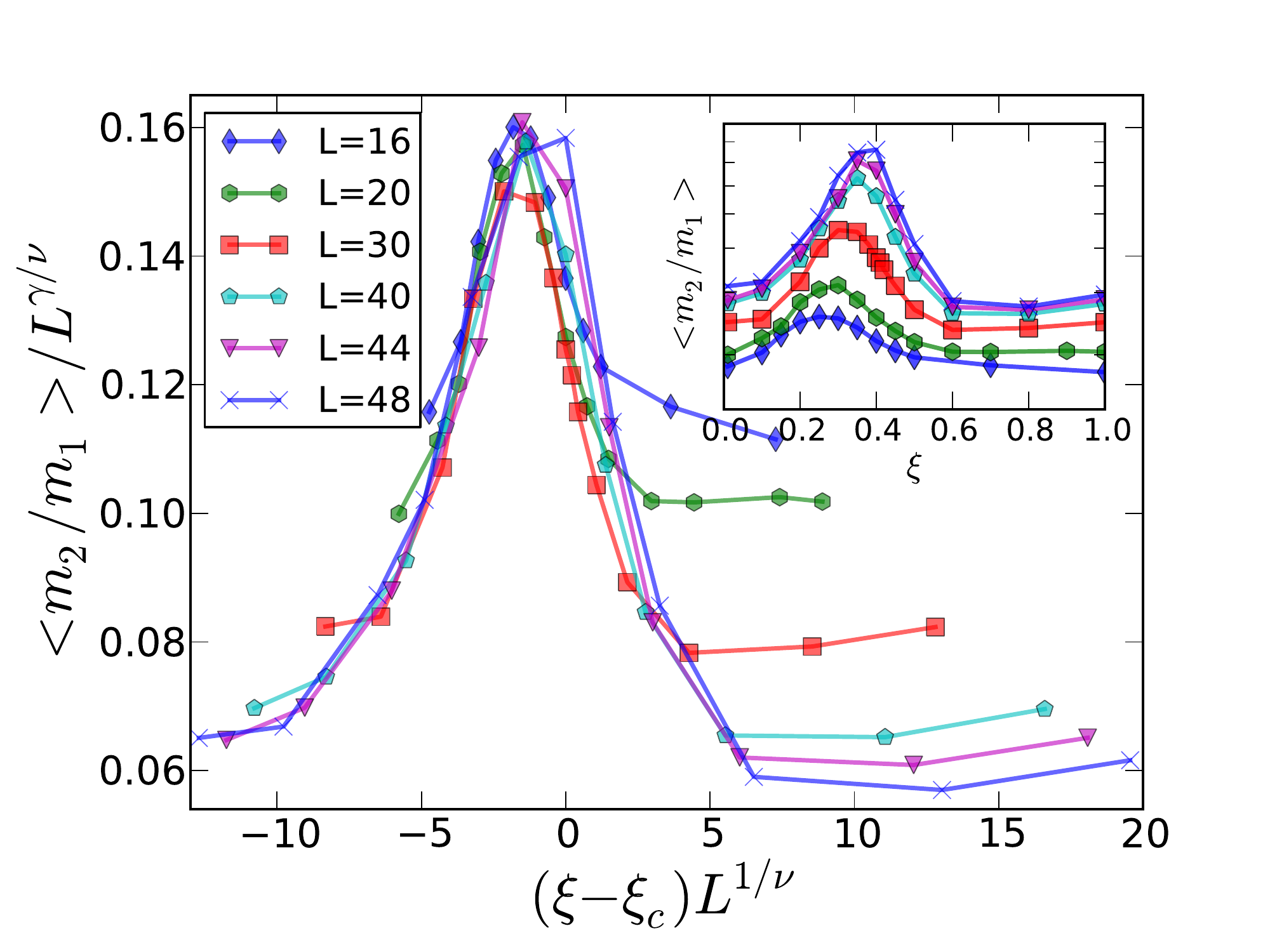}{Scaling plot of the average avalanche size $\left< m_2/m_1 \right>$. By rescaling the two axis with appropriate powers of $L$, all curves obtained for different $L$ values collapse on the top of each other for $\gamma / \nu = 1.$, $1 / \nu = 0.9$ and $\xi_c=0.4$. The inset shows the original data before rescaling.}
By increasing the system size $L$ the maximum of the average burst size $\left<\Delta\right>$ gets sharper, i.e.\ it becomes a higher and narrower peak. Based on the analogy to phase transitions, we tested the validity of the scaling form
\begin{eqnarray}
  \left<\Delta\right>(L,\xi) = L ^{\gamma/\nu}g((\xi-\xi_c)L^{1/\nu}) \quad,
  \label{eq:scale_averburst}
\end{eqnarray}
where $g(x)$ denotes a universal scaling function. It can be observed in Fig.\ \ref{eq:scale_averburst} that this scaling law allows for a data collapse of very good quality. The value of the exponents were obtained as $\gamma/\nu= 1$ and $1/\nu = 0.9$. It has to be emphasized that these values of $\xi_c$ and $1/\nu$ are consistent with the ones obtained from the scaling analysis of the largest burst $\left< \Delta_{max}\right>$.

\section{Discussion}
We carried out a theoretical investigation of the desiccation-induced fracture of a thin layer of an heterogeneous material. Previously, a large variety of experiments and computational modeling have shown that the crack patterns in such a system strongly depend on the amount of disorder of the material.  In the present study we focused on the temporal evolution of the fracture process by analyzing the statistics of avalanches of micro-fractures varying the amount of disorder. For this purpose a triangular lattice of spring elements was introduced where drying was captured by gradually decreasing the natural length of all springs. The quenched disorder of the material was represented by the 
random breaking thresholds of springs. The threshold values were uniformly distributed such that the amount of disorder could be controlled by varying the width of the distribution. 
The model was investigated by carrying out extensive computer simulations for several amounts of disorder in a broad range of system sizes.

As the most remarkable outcome, our study revealed that behind the formation of crack patterns induced by drying there is an interesting bursting activity of breakings. These bursts are sudden avalanches of fractures, induced by the breaking of a single material element. Simulations showed that the amount of disorder has a very strong effect on the statistics of avalanches: At low disorders, few large avalanches at the beginning of the drying create a large free surface in the system, and thereafter only small-sized avalanches can emerge.  The large avalanches are comparable to the size of the system and they distribute like a peak around a mean size, while the small ones show an exponential distribution of sizes. In the opposite limit of high disorder the entire breakup of the layer proceeds in small-sized bursts with an exponential distribution of sizes. Between the high and low disorder phases a sharp transition occurs at a critical amount of disorder of $\xi_c = 0.40+\pm 0.01$, where the avalanche size distribution becomes a power law.  The exponent of this power law is $\tau=2.6\pm 0.08$, in agreement with the mean-field value $\tau=5/2$ of the fiber bundle model.

To obtain a deeper understanding of this disorder-induced transition, we investigated the average size of the largest and second largest bursts as a function of disorder and performed a finite size scaling analysis of the results. Actually, the average value of the largest burst $\left< \Delta_{max}\right>$ can be identified as the order parameter, and a good quality finite-size scaling gives $\beta/\nu = 1.4$ and $1/\nu=1.0$ for the critical exponents. Similarly, the average ratio $\left< m_2/m_1\right>$ of the second $m_2$ and first moments $m_1$ of the avalanche size distribution shows a maximum at the same critical disorder $\xi_c=0.4$, and the finite-size scaling reaches the collapse of all curves for different system sizes $L$ with $\gamma / \nu = 1.$, $1 / \nu = 0.9$. These critical exponents characterize the transition. 

We note that in real systems the drying process might not be quasi-static, i.e.\ avalanches might be triggered by more than one micro-fractures, which also affects the statistic of avalanches. Recently, it has been shown analytically in the framwork of fiber bundles that compared to the quasi-static limit, the power law exponent of the burst size distribution gets higher when the system is driven with finite load increments \cite{PradhanHansenEtAlFailure09,hemmer_failure_2007}.  
A similar change of the burst size exponent can be expected when increasing the drying rate 
of the desiccation process. 

Desiccation induced cracking of pastes is a good candidate for a possible experimental realization of our theoretical findings. It has been shown recently in several experiments \cite{PhysRevE.60.6449,PhysRevE.74.045102} that pastes offer several opportunities to control the amount of disorder in the drying material: By varying the composition of the system (for instance by mixing two different materials such as polymers and clay, as in Ref.\ \cite{PhysRevE.60.6449}) a quenched structural disorder of local strength values can be introduced. Another possibility is to consider a single component paste, but varying the size distribution of particles in the colloidal suspension (as in \cite{PhysRevE.74.045102}). Both methods might be used to prepare samples with several amounts of quenched disorder, and high-speed imaging techniques could be used to monitor the avalanches by analyzing the temporal fluctuations of the crack pattern, as in \cite{wittel_fragmentation_2004,maloy_local_2006}. This experimental setup would contrast our results for the order parameter, the critical exponents and the phase transition itself.

\section*{Acknowledgments}
 We thank \href{http://www.colciencias.gov.co}{\emph{COLCIENCIAS}}
 (``Convocatoria Doctorados Nacionales 2008''),
 \href{http://www.ceiba.org.co}{\emph{Centro de Estudios
     Interdisciplinarios B{\'a}sicos y Aplicados en Complejidad- CeiBA
     - Complejidad}} and
 \href{http://www.unal.edu.co}{\emph{Universidad Nacional de
     Colombia}} for financial support. Most of the graphs were built
 using Matplotlib, \cite{Matplotlib}.  The work is supported by
 TAMOP-4.2.1/B-09/1/KONV-2010-0007 project. The project is implemented
 through the New Hungary Development Plan, co-financed by the European
 Social Fund and the European Regional Development Fund.  F.\ Kun
 acknowledges the support of OTKA K84157 and of the B\'olyai J\'anos
 foundation of the Hungarian Academy of Sciences.  This work was
 supported by the European Commissions by the Complexity-NET pilot
 project LOCAT.
\bibliography{VILLALOBOS-MUD}

\end{document}